\documentclass{PoS}

\title{Numerical Analysis of Discretized ${\cal N}=(2,2)$ SYM on Polyhedra}

\ShortTitle{Numerical Analysis of Discretized ${\cal N}=(2,2)$ SYM on Polyhedra}

\author{\speaker{Syo Kamata} \\
Physics Department and Center for Particle and Field Theory, Fudan University, \\
220 Handan Rd., Yangpu District, Shanghai 200433, China 
\footnote{Previous affiliation:  Hiyoshi Departments of Physics, and Research and Education Center for Natural Sciences, Keio University,  4-1-1 Hiyoshi, Yokohama, Kanagawa 223-8521, Japan}\\
E-mail: \email{skamata@rikkyo.ac.jp}}

\author{So Matsuura \\
  Hiyoshi Departments of Physics, and Research and Education Center for Natural Sciences, Keio University, 4-1-1 Hiyoshi, Yokohama, Kanagawa 223-8521, Japan\\
E-mail: \email{s.matsu@phys-h.keio.ac.jp}}

\author{Tatsuhiro Misumi \\
  Mathematical Science Course, Akita University, Akita 010-8502, Japan\\
  Research and Education Center for Natural Sciences, Keio University, \\
  4-1-1 Hiyoshi, Yokohama, Kanagawa 223-8521, Japan\\
E-mail: \email{misumi@phys.akita-u.ac.jp}}

\author{Kazutoshi Ohta \\
  Institute of Physics, Meiji Gakuin University, Yokohama 244-8539, Japan\\
E-mail: \email{kohta@law.meijigakuin.ac.jp}}

\abstract{

  We perform a numerical simulation of the two-dimensional ${\cal N}=(2,2)$ supersymmetric Yang-Mills (SYM) theory on the discretized curved space.
  The $U(1)_{A}$ anomaly of the continuum theory is maintained also in the discretized theory as an unbalance of the number of the fermions.
  In the process, we propose a new phase-quenched approximation, which we call the ``anomaly-phase-quenched (APQ) method", to make the partition function and observables well-defined by $U(1)_{A}$ phase cancellation.
  By adopting APQ method, we estimate the Ward-Takahashi identity for exact SUSY on lattice and clarify contribution of the pseudo zero-modes to the pfaffian phase.
}  

\FullConference{34th annual International Symposium on Lattice Field Theory\\
		24-30 July 2016\\
		University of Southampton, UK}

\newcommand{\be}{\begin{equation}}
\newcommand{\ee}{\end{equation}}
\newcommand{\bea}{\begin{eqnarray}}
\newcommand{\eea}{\end{eqnarray}}
\newcommand{\nl}{\nonumber \\}

\newcommand{\Tr}{{\rm Tr}}

\begin{document}

\section{Introduction}

Although the application of lattice formulation to supersymmetric (SUSY) theory is halfway, the lattice numerical analysis could lead to deeper understanding of SUSY nonperturbative phenomena \cite{lattice2016}. 
In this study, we perform numerical simulations for the two dimensional ${\cal N}=(2,2)$ SYM theory 
on curved background with a nontrivial topology.

In the continuum two dimensional ${\cal N}=(2,2)$ SYM, the $U(1)_{A}$ symmetry is anomalously broken 
due to the fermion measure in the partition function.
The $U(1)_{A}$ is expected to be also anomalous on the lattice and be hidden in the pfaffian obtained by integrating out fermions.
It is notable that, due to the anomaly, the partition function itself gets ill-defined.
In order to obtain reasonable and well-defined results, we propose a new phase-quenching method, which we call the {\it anomaly-phase-quenching}(APQ)\cite{Kamata:2016xmu}. Here, the Pfaffian phase associated with $U(1)_{A}$ anomaly is cancelled by inserting appropriate operators (called ``compensators") while the residual Pfaffian phase is ignored as with the standard phase-quenching procedure.
Since the latter Pfaffian phase is shown to have no influence on expectation values on the torus
we speculate this quenching procedure works.

We apply APQ to calculation of the Ward-Takahashi identity associated with exact SUSY on the lattice and obtain a result consistent with the analytical investigation. 
Moreover, we investigate the origin of $U(1)_{A}$ anomaly by looking into eigenstates of the Dirac operator and identify the pseudo zero-modes responsible for the $U(1)_{A}$-anomaly Pfaffian phase.

\section{Two dimensional SYM on discretized spacetime}
The $SU(N_{c})$ SYM action on a polyhedron has been constructed \cite{Matsuura:2014kha} based on the Sugino model\cite{Sugino:2003yb}.
In this theory, field components are defined on elements of a polyhedron, namely, on sites, links, and faces. A scalar SUSY $Q$ is preserved and it has nilpotency up to complexified gauge transformation, $Q^2 = \delta_{\phi}$.

The action is defined as
\bea
&& S_0 = Q\Xi \equiv Q \left[ \sum_{s=1}^{N_S} \alpha_{s} \Xi_{s} 
 + \sum_{l=1}^{N_L}  \alpha_{l} \Xi_{l} 
 + \sum_{f=1}^{N_F} \alpha_{f} \Xi_{f} 
 \right],
 \label{generalized Sugino} \\
&& \Xi_{s} = \frac{1}{2g^{2}}  \Tr \left[ \frac{1}{4} \eta_{s} [\Phi_{s},\bar{\Phi}_{s}] \right], \qquad
\Xi_{l} = \frac{1}{2g^{2}}  \Tr \left[ -i \lambda_{l} (U_{l} \bar{\Phi}_{{\rm tip}(l)} U^{-1}_{l} - \bar{\Phi}_{{\rm org}(l)}) \right] , \nl
&& \Xi_{f} = \frac{1}{2g^{2}}  \Tr \left[  \chi_{f} ( Y_{f}- i \beta_{f} \Omega(U_{f})  ) \right],
\eea
where $g^2$ is a coupling constant, $\alpha_{s,l,f}$ and $\beta_{f}$ are arbitrary parameters, and $N_{S},N_{L},N_{F}$ are the number of sites, links, and faces, respectively.
In this study, we choose $\alpha_{s,l,f}$ and $\beta_{f}$ as unity.
The complex scalar fields $\Phi_{s}$ and $\bar{\Phi}_{s}$ are defined on sites, and the link fields $U_{l}$ are on oriented links, and the auxiliary fields $Y_{f}$ are on faces.
In addition, fermions $\eta_{s},\lambda_{l},\chi_{f}$ also live on sites, links, and faces, respectively.
The symbols ${\rm org}(l)$ and ${\rm tip}(l)$ denote the origin and tip of the link $l$, respectively.
$U_{f}$ is a Wilson loop along the edges surrounding the face $f$.
$\Phi_{f}$ is a scalar field on a representative site of the face.
To eliminate unphysical degenerate vacua
\cite{Matsuura:2014pua}, we adopted 
\bea
&& \Omega(U_{f}) = 
\frac{1}{m} \left[ {\mathcal S}^{-1}(U^{m}_{f}) {\mathcal C}(U^{m}_{f})
+ {\mathcal C}(U^{m}_{f}) {\mathcal S}^{-1}(U^{m}_{f})\right],  
\quad \left(m \ge \frac{N_c}{4} \right), \label{moment map}
\eea
where ${\mathcal S}(U_{f}) = -i ( U_{f}-U^{-1}_{f} )$ and ${\mathcal C}(U_{f}) = U_{f}+U^{-1}_{f}$.
The SUSY transformation is defined as
\bea
&& Q \Phi_{s} = 0, \quad Q \bar{\Phi}_{s} = \eta_{s}, \quad Q U_{l} = i \lambda_{l} U_{l}, \quad Q Y_{f} = [\Phi_{f},\chi_{f}],\nl
&& Q \eta_{s} = i [\Phi_{s} , \bar{\Phi}_{s}], \quad Q \lambda_{l} = i (U_{l} \Phi_{{\rm tip}(l)} U^{-1}_{l} - \Phi_{{\rm org}(l)} + \lambda_{l} \lambda_{l}), \quad Q \chi_{f} = Y_{f}. 
\eea
Since the action has a $Q$-exact expression, it is trivially invariant under the SUSY transformation. 
In addition, the transformation is closed on each elements (sites, links and faces) of the polyhedron, hence the action is also $Q$-invariant within each of them.
A SUSY breaking mass term is needed to control the flat direction of the scalar fields, which is introduced as
\be
S_{\mu} = \frac{\mu^2}{2} \sum_{s} \, \Tr (\Phi_{s} \bar{\Phi}_{s}),
\ee
where $\mu^2$ is a mass parameter.

\section{$U(1)_{A}$ anomaly and the pfaffian phase}
We now discuss the $U(1)_{A}$ anomaly and its relation with a pfaffian phase.
The $U(1)_A$ transformation in the continuum SYM theory is defined as
\bea
&& A_{\mu} \rightarrow A_{\mu}, \quad \ \ \ \ \Phi \rightarrow e^{2i \theta}\Phi, \quad \bar{\Phi} \rightarrow e^{-2i \theta} \bar{\Phi}, \quad Y \rightarrow Y , \nl
&&
 \lambda_{\mu} \rightarrow e^{i\theta} \lambda_{\mu}, \quad \eta \rightarrow e^{-i\theta} \eta, \quad \chi \rightarrow e^{-i \theta} \chi,
\eea
which is anomalous as mentioned above.
The $U(1)_{A}$ is also anomalous on the lattice.
Since the anomaly arises from the fermion measure, 
a Pfaffian phase of the Dirac operator obtained after integrating out fermions is responsible for it.
To clarify it, we define the partition function on the lattice as
\be
I = \int {\cal D}\vec{B} {\cal D}\vec{F} \, e^{-S_{0,b}-S_{0,f}-S_{\mu}} \,=\, \int {\cal D}\vec{B} \, {\rm Pf}(D) \,e^{-S_{b}},
\ee
with $S_{b}=S_{0,b}+S_{\mu}$, where $S_{0,b}$ and $S_{0,f}$ are the bosonic part and fermionic part of the action (\ref{generalized Sugino}), ${\cal D}\vec{B}$ and ${\cal D}\vec{F}$ are the integrate measure of bosons and fermions, respectively, and ${\rm Pf}(D)$ is the pfaffian of the Dirac operator.
The partition function is not $U(1)_{A}$ neutral due to the fermion measure: the measure of the partition function have the following net $U(1)_{A}$ charge,
\be
[{\cal D} \vec{B} {\cal D}\vec{F}]_{A} = (N_{c}^2-1) \chi_{h}, 
\label{U1charge}
\ee
where $\chi_{h}$ is the Euler characteristics.
After integrating out the fermions, the Pfaffian has a nontrivial phase
\be
 {\rm Pf}(D) =  | {\rm Pf}(D) | e^{i \theta_{\rm pf}} , \quad \theta_{\rm pf} = \theta_{A}+ \theta,
\ee
where $\theta_{A}$ is the $U(1)_{A}$-anomaly-induced phase, which will be defined later, and $\theta$ is a residual phase apart from the $U(1)_{A}$ phase.

As Eq.~(\ref{U1charge}) indicates, the anomaly directly reflects the number of fermionic degrees of freedom on the sites, links and faces: Nonzero Euler characteristics means that sum of the number of fermion degrees freedom on sites and faces is different from those on links.
Unless the Euler characteristics is zero, the partition function is not $U(1)_{A}$ neutral and becomes ill-defined. We thus need to define $U(1)_{A}$ neutral definitions of partition function and expectation values.


For our purpose, we consider the following procedure:

Firstly, we introduce an operator ${\cal A}$ which satisfies gauge invariance, exact SUSY invariance, and $[{\cal A}]_{A} = -(N_{c}^2-1) \chi_{h}$. It is clear that this operator, which we call the ``compensator", has a $U(1)_{A}$ charge canceling out that from the fermion measure.
We define the $U(1)_A$ anomaly-induced phase $\theta_A$ through $A=|A|e^{-i \theta_A}$.
For later convenience, we introduce two types of compensators,
\bea
&& {\mathcal A}_{\rm tr}
\,=\, \frac{1}{N_S} \sum_{s=1}^{N_S} 
\left( 
 \frac{1}{N_c} {\rm Tr}\left( \Phi_{s} \right)^2 
\right)^{-\frac{N_c^2-1}{4}\chi_h}, 
\label{tr_comp}
\nl
&& {\mathcal A}_{\rm IZ}
\,=\, \frac{1}{N_l} \sum_{l=1}^{N_l} 
\left( \frac{1}{N_c} {\rm Tr}\left( 2\Phi_{{\rm org}(l)} U_{l}\Phi_{{\rm tip}(l)} U_{l}^{\dag}+\lambda_{l}\lambda_{l} (U_{l}\Phi_{{\rm tip}(l)} U^{\dag}_{l}+\Phi_{{\rm org}(l)})\right)
\right)^{-{{N_{c}^{2}-1}\over{4}}\chi_{h}}.
\label{IZ_comp}
\eea
Although both operators have the same $U(1)_{A}$ charge, the second operator includes not only the scalar field but also fields on links.
Since the $\bar{\Phi}_{s}$ is related to $\eta_{s}$ through $Q$ and the anomaly is caused by remained zeromodes of $\eta_s$ and $\chi_f$, the first operator might be effective for $h=0$. 
In the same sense, it is better to employ the second operator for $h=2$ because the anomaly is induced by those of $\lambda_{l}$.

Secondly, we introduce the new quenching procedure for expectation values of operators $\mathcal O$ as 
\bea
\langle {\cal O} \rangle ^{\hat{q}} \equiv \langle {\cal O} e^{i \theta_{A}} \rangle ^{q}, \quad \langle {\cal O} \rangle^{q} \equiv \frac{1}{Z_{q}} \int {\cal D} \vec{B} \, {\cal O} |{\rm Pf}(D)| \, e^{-S_{b}}.
\eea
where $Z_{q}$ is the quenched partition function.
this quenching procedure $\langle {\cal O} \rangle ^{\hat{q}}$ ignore only the Pfaffian phase unrelated to the $U(1)_{A}$ anomaly.
This residual Pfaffian phase is shown to have no influence on expectation values on the torus in the literature, and we expect this procedure works on generic backgrounds.
We call it {\it Anomaly-phase-quenching}(APQ).

Thirdly, we combine the above two methodologies:
We insert the compensator into the expectation values of observables with applying APQ method.
Here, the anomaly-induced Pfaffian phase is cancelled by the compensator while the residual phase can be ignored by APQ.

We end up with a ``good" definition of the expectation values in the present discretized model.
We call the whole procedure the ``APQ method".
If we obtain a reasonable result on the expectation value by this method, it means not only validity of the model but also validity of our quenching method.
In the present study, we numerically calculate the Ward-Takahashi(WT) identity for the exact supersymmetry using the APQ method.
We estimate the following identities,
\bea
&& \langle \tilde{S}_{b}{\mathcal A}_{\rm tr} \rangle\, 
+\,  {\mu^{2}\over{2}} \sum_{s} \langle \Xi {\rm Tr}(\Phi_{s} \eta_{s}){\mathcal A}_{\rm tr}  \rangle 
- {N_{c}^{2}-1 \over{2}}(N_S+N_L)\langle {\mathcal A}_{\rm tr}  \rangle =0, \,  
\label{PCSC2-boson}
\eea
for $h=0$ and
\bea
&&\langle \tilde{S}_{b}{\mathcal A}_{\rm IZ} \rangle\, 
+\,  {\mu^{2}\over{2}} \sum_{s} \langle \Xi {\rm Tr}(\Phi_{s} \eta_{s}){\mathcal A}_{\rm IZ}  \rangle 
- {N_{c}^{2}-1 \over{2}}(N_S+N_L)\langle {\mathcal A}_{\rm IZ}  \rangle \nonumber \\
&&\hspace{1cm}-\frac{N_c^2-1}{4}\chi_h 
\Bigl\langle \frac{1}{N_L}  \sum_{l=1}^{N_L}
\frac{1}{N_c}{\rm Tr}
\left( 
\lambda_l \lambda_l 
\left(U_l \Phi_{{\rm tip}(l)} U_l^\dagger +\Phi_{{\rm org}(l)}\right)
\right) \nonumber \\
&&\hspace{1cm}\times
\left( \frac{1}{N_c} {\rm Tr} \left(
2\Phi_{{\rm org}(l)} U_l \Phi_{{\rm tip}(l)} U_l^\dagger
+\lambda_l \lambda_l 
\left(U_l \Phi_{{\rm tip}(l)} U_l^\dagger +\Phi_{{\rm org}(l)} \right)
\right)\right)^{-\frac{N_c^2-1}{4} \chi_{h}-1}
\Bigr\rangle
=0,\,  
\label{PCSC2-IZ}
\eea
for $h=2$ background, where $\tilde{S}_{b}$ is the bosonic action after integrating out the auxiliary field $Y$.
We numerically calculate these WT identities based on the APQ method, where
the expectation values $\langle... \rangle$ in the above equations are replaced by $\langle... \rangle^{\hat q}$.

\begin{figure}[thbp]
  \begin{minipage}{0.65\textwidth}
    \begin{center}
      \makeatletter
      \def\@captype{table}
      \makeatother
      \begin{scriptsize}
        \begin{tabular}{|cc|ccccccc|} \hline
           $h$ & $\chi_h$ & geometry & $N_S$ & $N_L$ &  $N_F$ & shape of face & $a$ & \\ \hline \hline
           $0$ & 2 & tetra & 4 & 6 & 4 & T & 0.7598 & \\
            & & octa & 6 & 12 & 8 & T & 0.5373 &  \\
            & & cube & 8 & 12 & 6 & S & 0.4082 & \\
            & & icosa & 12 & 30 & 20 & T & 0.3398 & \\ 
            & & dodeca & 20 & 30 & 12 & P & 0.2201 & \\ \hline
           $1$ & 0  & $3 \times 3$ reg.lat. & 9 & 18 & 9 & S & 0.3333 & \\ 
                        &    & $4 \times 4$ reg.lat. & 16 & 32 & 16 & S & 0.2500 & \\
                        &    & $5 \times 5$ reg.lat. & 25 & 50 & 25 & S & 0.2000 & \\ \hline
           $2$ & -2  & Right fig. &14 &32& 16 & S & 0.2500 & \\ \hline
        \end{tabular}
      \end{scriptsize}
        \end{center}    
    \label{tab:sim_param}
  \end{minipage}
  \begin{minipage}{0.35\hsize}
    \begin{center}
      \includegraphics[clip,width=35mm]{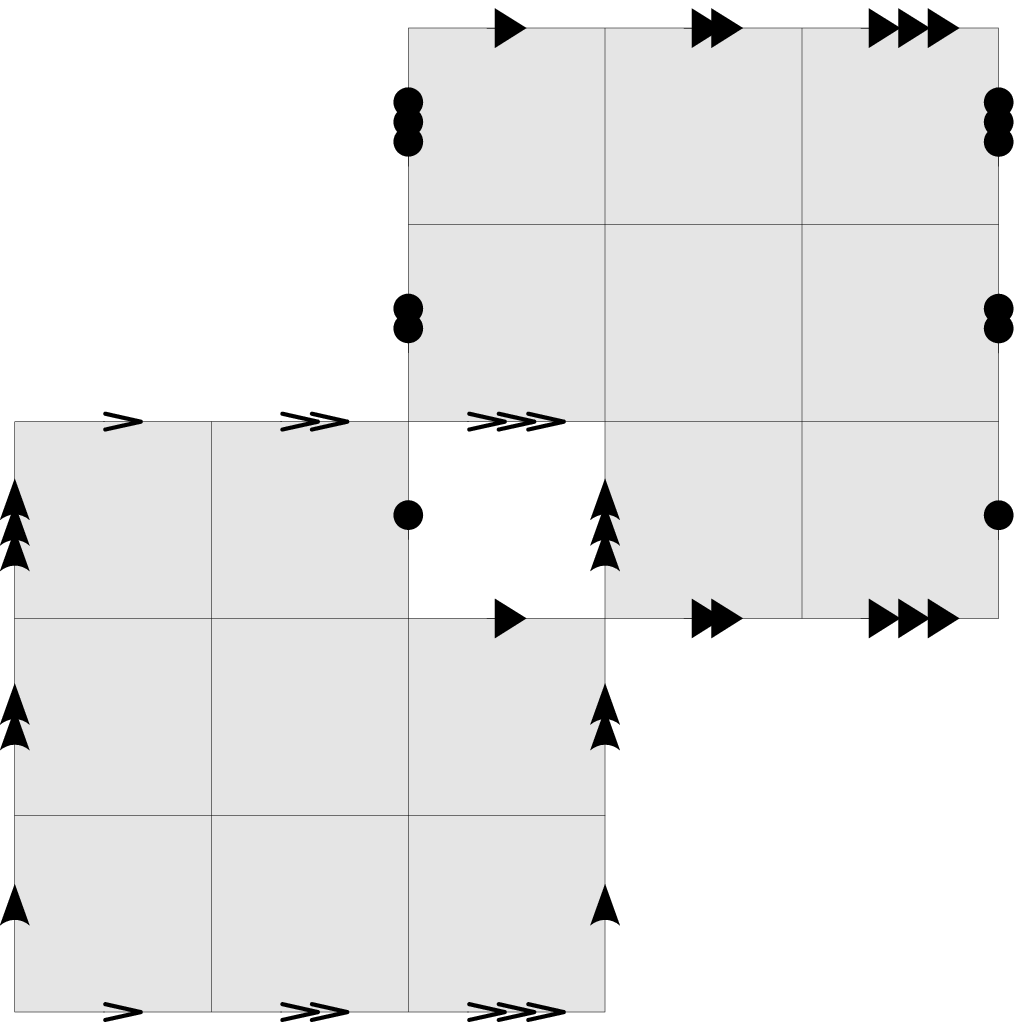} 
    \end{center}
    \label{fig:genus2}
  \end{minipage}
        \makeatletter
      \def\@captype{table}
      \makeatother
      \caption{\small List of polyhedra and those topologies used in our simulations. The symbols in the fifth column T, S, P express ``triangle'', ``square'' and ``pentagon'', respectively.
        The right figure shows the development view 
        used in our simulation for $h=2$.}
\end{figure}

\section{Numerical simulation}
We perform Monte Carlo simulations of the SYM theory on background with $h=0,1,2$ topologies. 
In this simulation, we not only confirm the WT identity of the exact SUSY which depends on background topology but also verify the validity of the APQ method. 
Our setup for the numerical simulations is presented in Tab.1.

\begin{figure}[htbp]
  \begin{center}
    \begin{minipage}{0.32\hsize}
      \begin{center}
        \includegraphics[clip,width=52mm ]{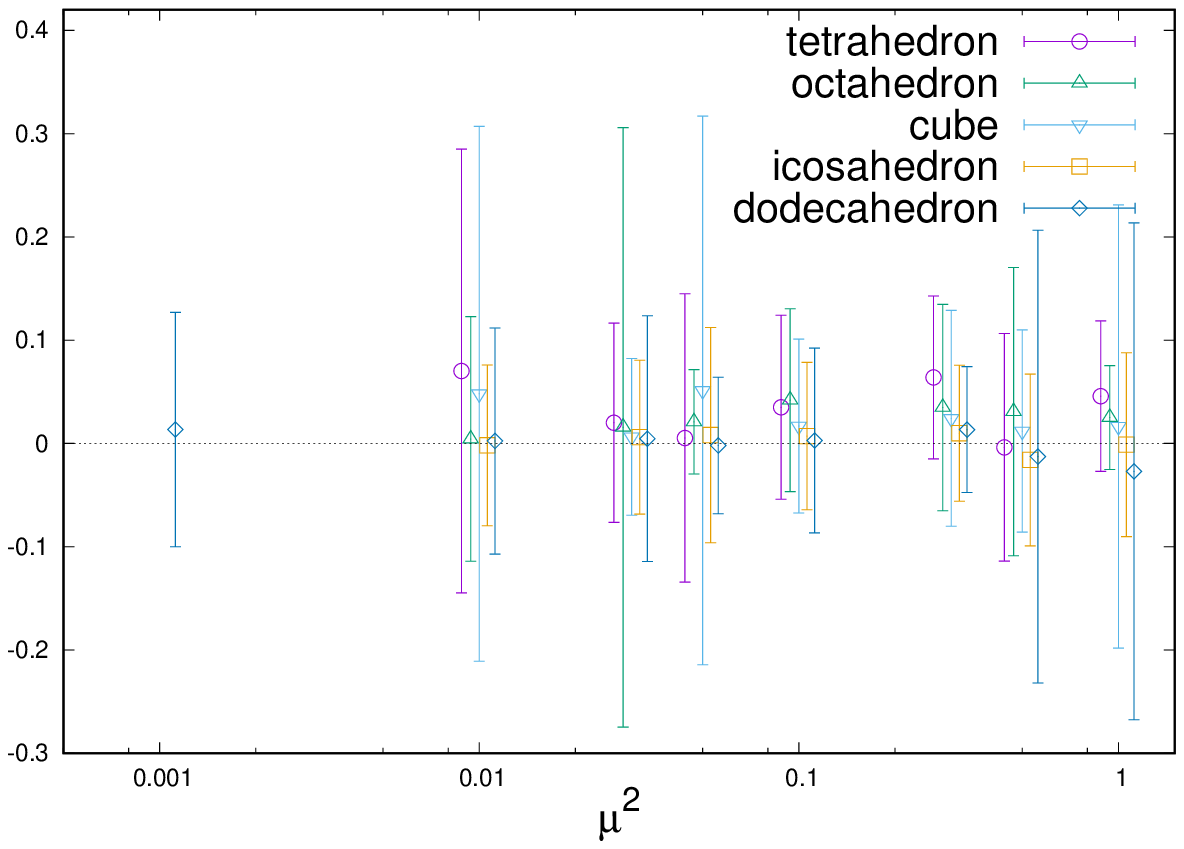} \\
                        {\scriptsize (1) WT identity for $h=0$}
      \end{center}   
    \end{minipage}
    \begin{minipage}{0.32\hsize}
      \begin{center}
        \includegraphics[clip,width=52mm]{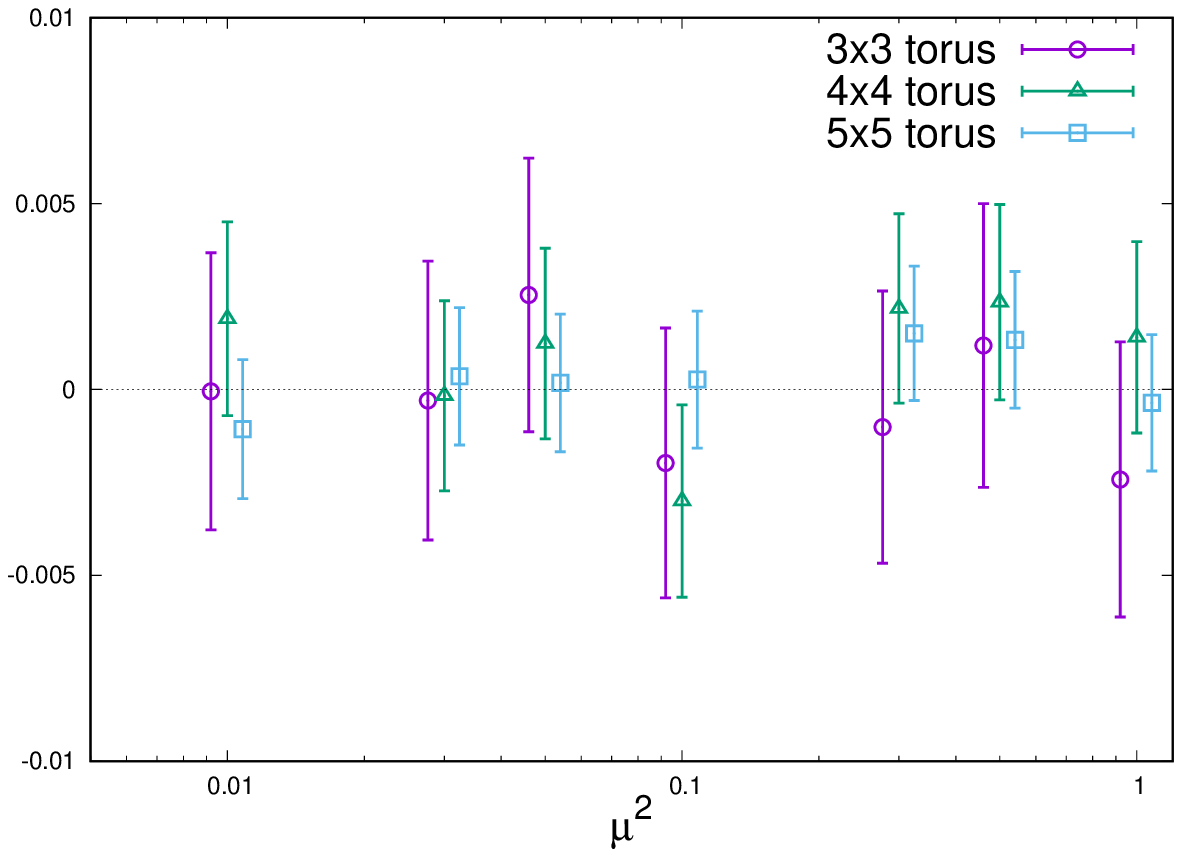}\\
                        {\scriptsize (2) WT identity for $h=1$}
      \end{center}
    \end{minipage} 
    \begin{minipage}{0.32\hsize}
      \begin{center}
        \includegraphics[clip,width=52mm]{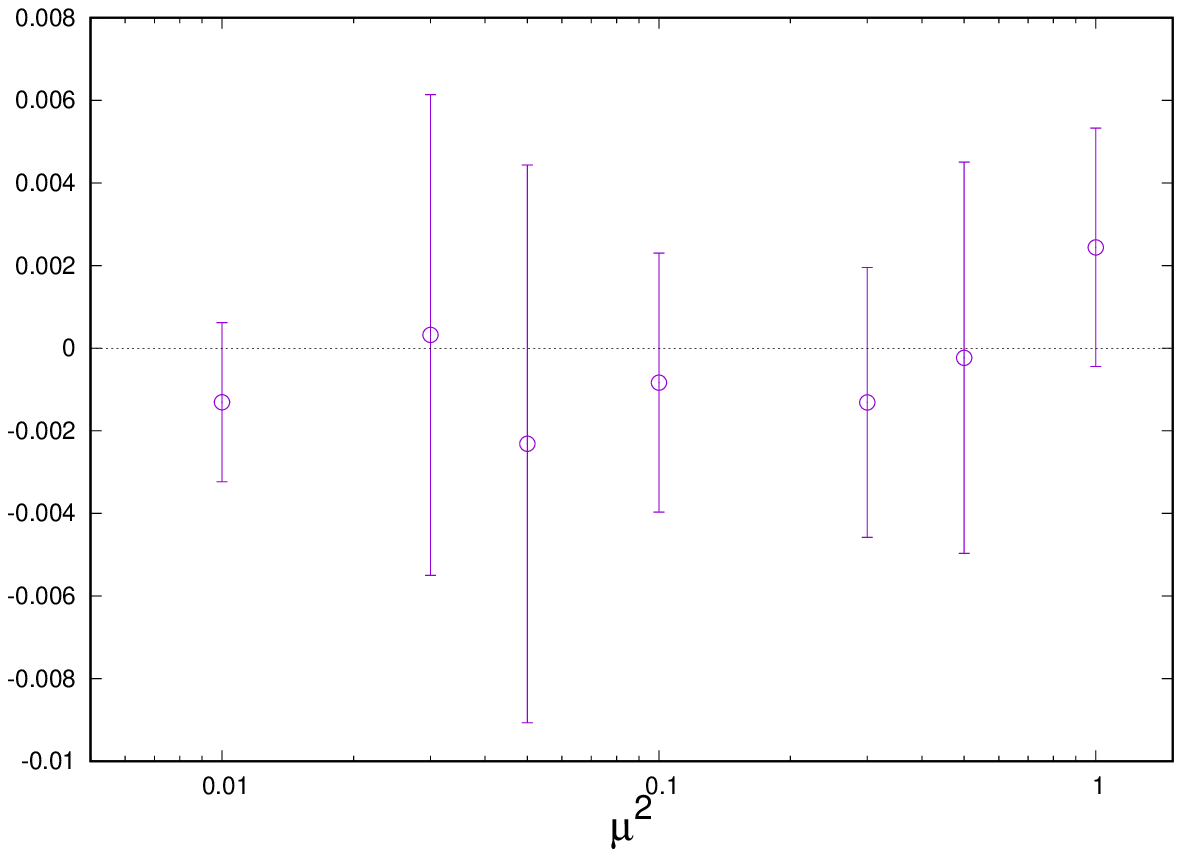}\\
                        {\scriptsize (3) WT identity for $h=2$}
      \end{center}
    \end{minipage} 
    \caption{\small
      The left hand side of the WT identities
      (3.7) (the panels (1) and (2))
      and 
      (3.8) (the panel (3))
      in the anomaly-phase-quenched approximation 
      normalized by 
      $\frac{1}{2}(N_c^2-1)(N_S+N_L)\langle {\cal A}\rangle^{\hat{q}}$ against to $\mu^2$ 
      for $h=0$ (left), $h=1$ (middle) and $h=2$ (right). 
      We have used the compensator ${\cal A}_{\rm tr}$ for $h=0$ and
      ${\cal A}_{\rm IZ}$ for $h=2$ while we have set ${\cal A}=1$ for $h=1$ since we do not need 
      the compensator when $h=1$. 
    }    
    \label{fig:PCSC}
  \end{center}
\end{figure}

The Fig.\ref{fig:PCSC} shows numerical results of the left-hand sides of the WT identities
(\ref{PCSC2-boson}) and (\ref{PCSC2-IZ}).
As definition of the compensators, we used eq.(\ref{tr_comp}) 
for $h=0$ and eq.(\ref{IZ_comp}) for $h=2$.
The numerical results have good agreement with the theoretical predictions ${\rm l.h.s.}=0$ 
within error bars.
Note that this numerical check is nontrivial, because the WT identity is not satisfied if the compensate operators is not inserted.
Hence, this result also means that the APQ method does work.

\begin{figure}[htbp]
  \begin{center}
    \begin{minipage}{0.42\hsize}
      \begin{center}
        \includegraphics[clip, width=65mm]{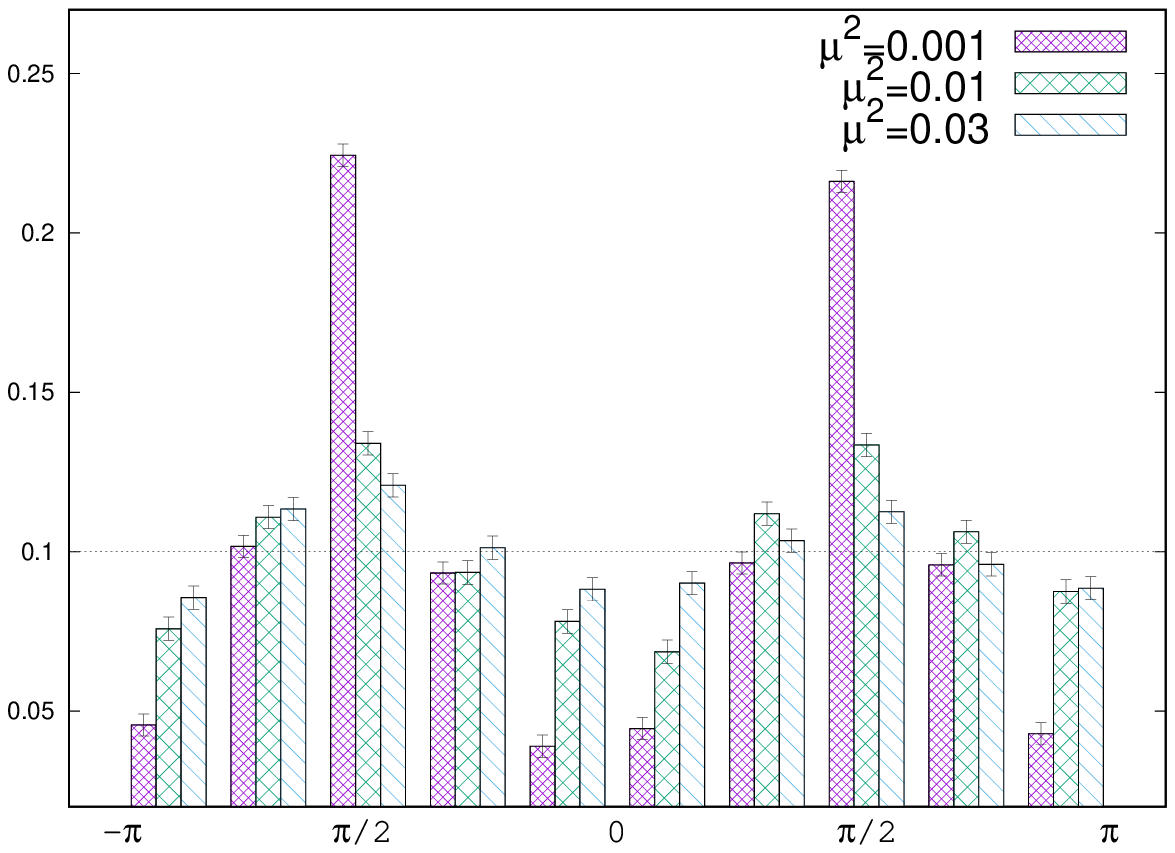} \\
                        {\scriptsize (1) dodecahedron}
      \end{center}
    \end{minipage} 
    \begin{minipage}{0.42\hsize}
      \begin{center}
        \includegraphics[clip, width=65mm]{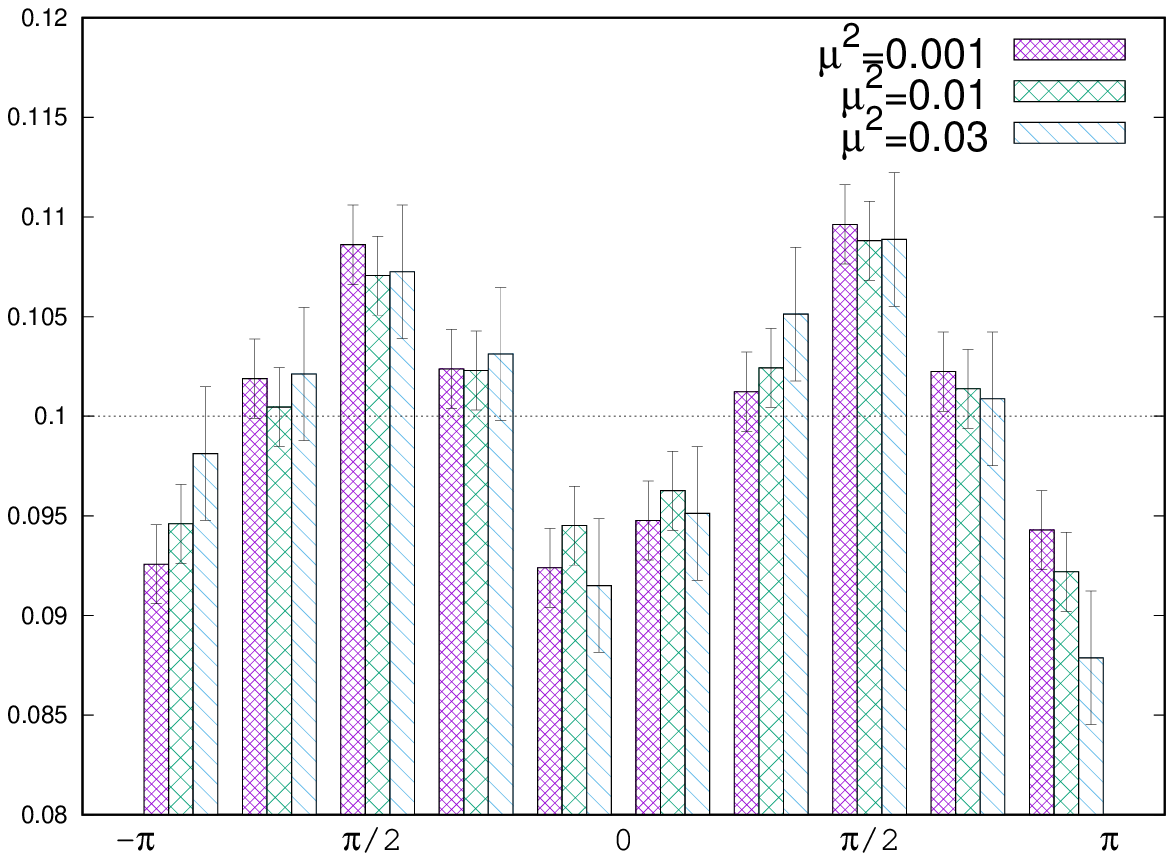} \\
            {\scriptsize (2) double torus}
      \end{center}
      \end{minipage} 
    \caption{\small The histogram of the phase of ${\rm Pf}(D){\cal A}_{\rm tr}$ 
      for the dodecahedron $h=0$ (left) and ${\rm Pf}(D){\cal A}_{\rm IZ}$  for $h=2$ (right). 
      The mass parameters are $\mu^2=0.01$, $0.1$ and $0.03$.}
    \label{fig:Apfaffian_phase}
  \end{center}
\end{figure}

The phase histogram obtained by $ {\rm Pf}(D) {\mathcal A} = | {\rm Pf}(D) {\mathcal A} | e^{i \theta} $ for $h=0$ and $h=2$ is shown in Fig.\ref{fig:Apfaffian_phase}.
This figure shows that two peaks appear around $\pm \pi/2$ and the peaks become sharper as taking the boson mass smaller.
The existence of the peaks means that the sign problem due to the $U(1)_{A}$ anomaly vanishes in the APQ method.
In addition, the validity of quenching the residual phase is shown by the result of the WT identity obtained by the APQ method.

\begin{figure}[htbp]
  \begin{center}
      \begin{minipage}{0.42\hsize}
        \begin{center}
          \includegraphics[clip, width=65mm]{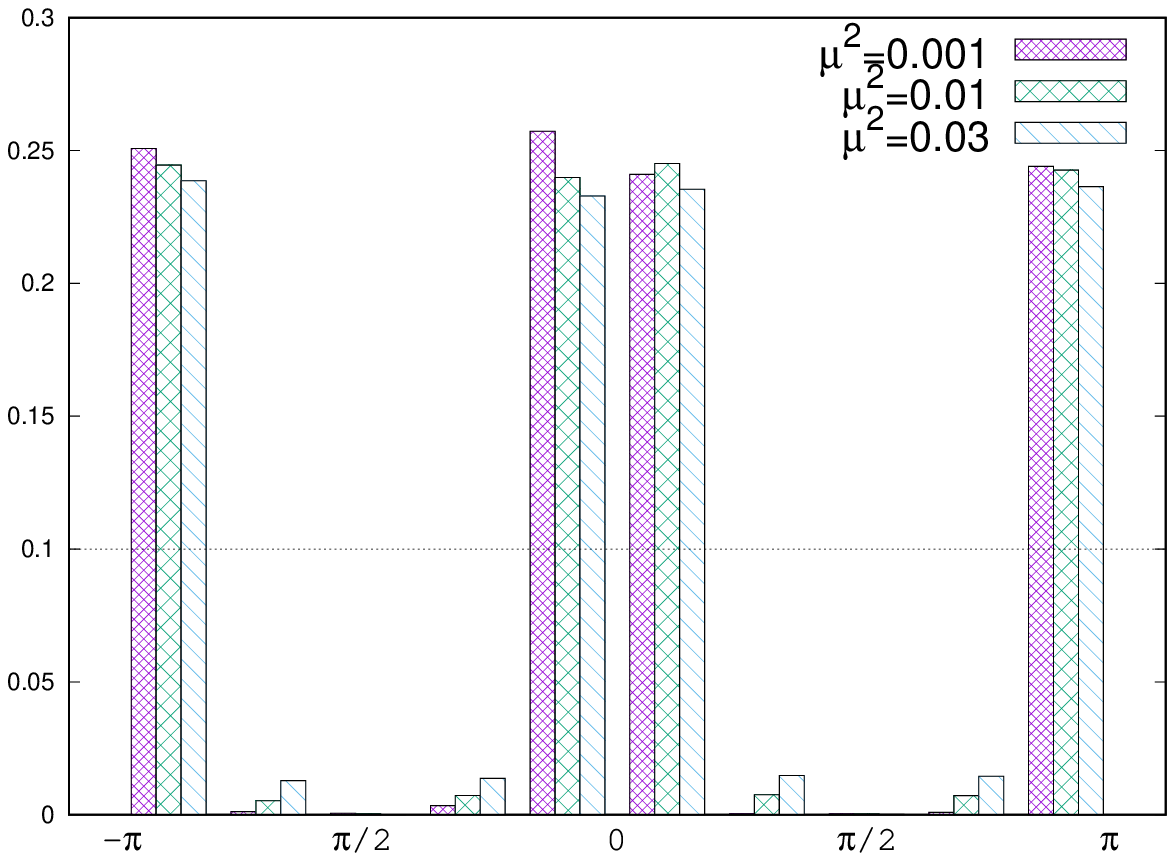} \\
          {\scriptsize (1) dodecahedron, $\mu^2=0.01,0.03,0.05$}
        \end{center}
      \end{minipage} 
            \begin{minipage}{0.42\hsize}
        \begin{center}
          \includegraphics[clip, width=65mm]{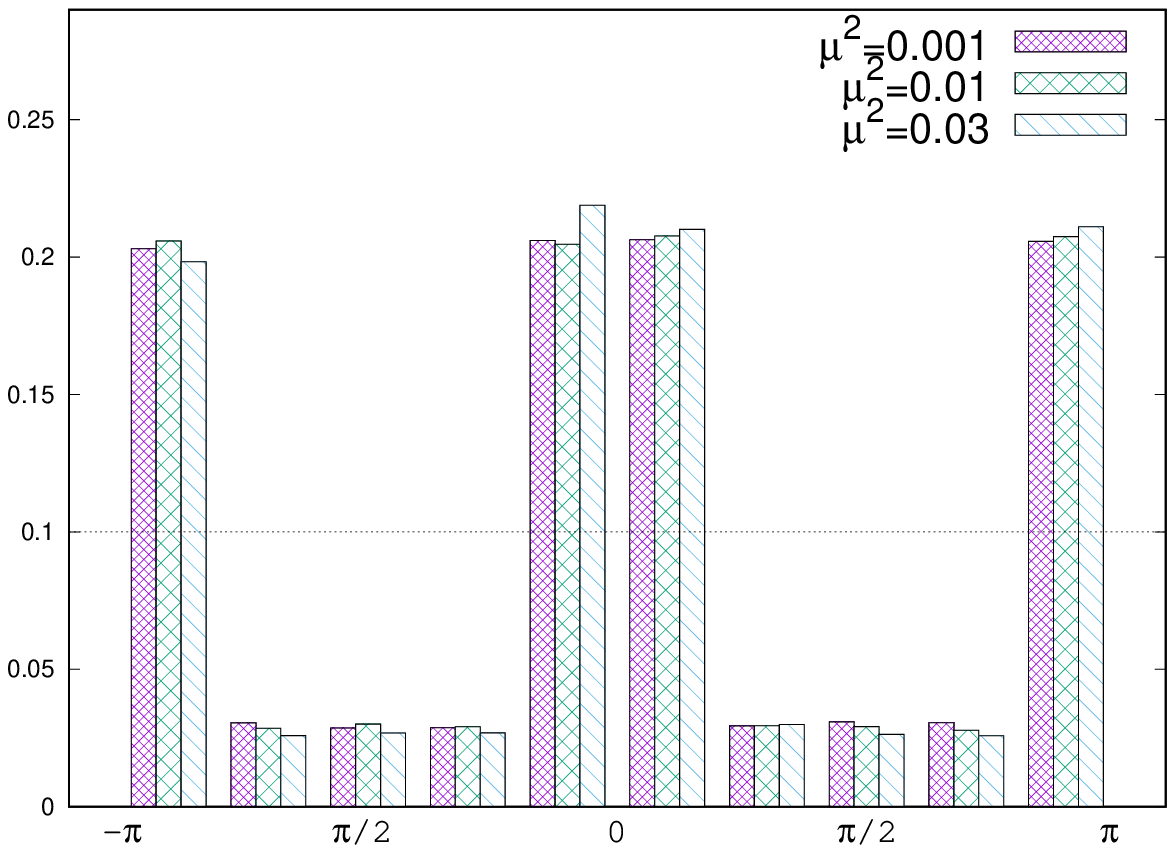} \\
             {\scriptsize (2) double torus, $\mu^2=0.001,0.01,0.03$}
        \end{center}
      \end{minipage} 
      \caption{\small This histogram of the phase of the subtracted Pfaffian ${\rm Pf}'(D)$
      for the dodecahedron (left) and the double torus (right). 
      }
    \label{fig:pfaffian_sub}
  \end{center}
\end{figure}

We also identify the pseudo zeromodes responsible for the $U(1)_{A}$ anomaly and subtract these contributions from the original pfaffian phase.
The histogram of the phase of this subtracted pfaffian ${\rm Pf}'(D)$ is presented 
in Fig.\ref{fig:pfaffian_sub}.
The result has quite sharp peaks around $\pm \pi$ both for $h=0$ and for $h=2$.
Hence, this fact shows that the anomaly-induced sign problem originates in the pseudo zeromodes and vanishes by removing those contribution.


\section{Summary and outlook}
We performed the numerical simulation of the ${\cal N}$=(2,2) SYM on discretized spacetime with a nontrivial topology.
In the theory, the $U(1)_{A}$ symmetry is generically broken by quantum anomaly and the partition function gets ill-defined.
In order to make the partition function well-defined and to obtain reasonable results, we introduced the anomaly-phase-quenched(APQ) method and numerically calculated the Ward-Takahashi(WT) identity associated with the exact SUSY.
Our results depending on topology of the background are consistent with the theoretical prediction, 
and the validity of the APQ method is also verified.
We investigated contribution of pseudo zeromodes to the pfaffian phase.
The phase has peaks, and the sign problem due to $U(1)_A$ anomaly vanishes by inserting the compensators.
Moreover, they become quite sharp after subtracting the pseudo zeromodes from the original phase.
This fact shows that the anomaly-induced sign problem and the $U(1)_{A}$ anomaly originate in the pseudo zeromodes and vanish by removing those contributions.

The construction of the other two dimensional gauge theories such as ${\cal N}=(4,4)$ and ${\cal N}=(8,8)$ is available in the similar way.
Another interesting issue is to construct matter coupled theories on a polyhedron.
By adding chiral multiplets to the discretized SYM theory, the obtained theory has a richer structure than the SYM theory.
It will be interesting to understand how it happens in the discretized theory and is now going.

\section*{Acknowledgements}
The work of S.M., T.M. and K.O. was supported in part by 
Grant-in-Aid for Scientific Research (C) 15K05060, 
Grant-in-Aid for Young Scientists (B) 16K17677,
and JSPS KAKENHI Grant Number JP26400256, respectively.
S.K. is supported by the Advanced Science Measurement Research Center at Rikkyo University.
This work is also supported by
MEXT-Supported Program for the Strategic Research Foundation
at Private Universities ``Topological Science'' (Grant No. S1511006).

\end{document}